%Paper: gr-qc/9507048
%From: amos@phys1.technion.ac.il (amos ouri)
%Date: Tue, 25 Jul 1995 11:24:04 -0100

%%%%%%%%%%%%%%%%%%%%%%%%%%%%%%%%%%%%%%%%%%%%%%%%%%%%%%%%%%%%%%%%%%%%%%
%%%%%%%%%%% Plain TeX file %%%%%%%%%%%%%%%%%%%%%%%%%%%%%%%%%%%%%%%%%%%
%%%%%%%%%%%%%%%%%%%%%%%%%%%%%%%%%%%%%%%%%%%%%%%%%%%%%%%%%%%%%%%%%%%%%%
%
\centerline {\bf RADIATIVE EVOLUTION OF ORBITS AROUND}
\vskip .4cm
\centerline {\bf A KERR BLACK HOLE}

\vskip 2cm

\centerline {Amos Ori}
\vskip 1cm
\centerline {Department of Physics, Technion -
 Israel Institute of Technology,}
\vskip 0.3cm
\centerline {Haifa, 32000, Israel}
\vskip 1cm
\centerline {Phys. Lett. A. {\bf 202}, 347 (3 July 1995)}

\vskip 3cm
\centerline {\bf Abstract}
\vskip 1cm
We propose a simple approach for the radiative evolution of
generic orbits around a Kerr black hole. For a scalar-field, we
recover the standard results for the evolution of the energy $E$
and the azimuthal angular momentum  $L_z$ . In addition, our
method provides a closed expression for the evolution of the
Carter constant $Q$.

\vskip 3cm
Recently, the prospects of a direct observation of gravitational
radiation led to a new interest in the old-standing problem of
radiation back-reaction. The most promising radiation sources for
the proposed gravitational-waves detector LIGO (as well as LISA
[1]) are coalescing binary systems. [2] In such a binary system,
energy and angular momentum are carried away by the emitted
gravitational radiation. As a consequence, the system becomes more
and more bound, resulting in a steady grow in the orbital
frequency and energy-loss rate. Eventually, the binaries will
coalless, releasing a strong burst of gravitational radiation. [2]
The prospects of observing such bursts raises the need for a
theoretical prediction of the evolution of such binary systems. If
one of the binary objects is a black hole, then curvature effects
could be important. This motivates one to understand the
radiation-reaction in curved spacetime.

	Consider the situation of a compact object with mass  $\mu $
orbiting a super-massive Kerr black hole with mass $M>>\mu $  and
angular momentum $aM$ (we are using general-relativistic units
C=G=1). This situation is particularly important for the proposed
space-based gravitational-waves detector LISA, [1,3] but is also
interesting in its own right. The problem of radiative evolution
of orbits around a Kerr black hole was treated by several authors
[4, 5], but so far with partial success only. The orbits in Kerr
geometry are characterized by three constants of the motion: the
energy $E$ , the azimuthal angular momentum $L_z$ , and the Carter
constant $Q$ . [6] Within the adiabatic approximation [3] that we
use here, the problem of radiative evolution is reduced to that of
calculating $\dot E$  , $\dot L_z$  and $\dot Q$  , where the
overdot denotes the evolution rate (in terms of external time
$t$). So far, expressions have been given for $\dot E$   and $\dot
L_z$   , based on the rate that energy and angular-momentum are
carried away to infinity and into the black hole [4,5,7]. However,
$\dot Q$  is still unknown. [8]

	Although we are primarily interested in the backreaction of
gravitational radiation, it would be convenient to generalize the
discussion and to consider electromagnetic and scalar-field
radiation as well. We shall thus assume that the particle carries
a "generalized charge" $q$ , which may be either a scalar,
electromagnetic, or gravitational charge (we demand $q<<M$; in the
gravitational case, $q\equiv \mu $ ). In the discussion below we
shall mainly refer to scalar and electromagnetic radiation (the
explicit calculations in this Letter will be restricted to the
scalar-field case) ; However, our approach can be extended to
gravitational radiation as well [9].

	Our main goal here is to propose a scheme for the calculation
of $\dot Q$  (as well as $\dot E$   and $\dot L_z$   ) for generic
orbits in Kerr spacetime, by a direct calculation of the
radiation-backreaction force - based on the pure retarded
potential. An analogous direct scheme for radiative evolution of
$E$ and $L_z$ has already been proposed by Gal'tsov [4] , but that
scheme is based on the so-called "radiative potential" (the "half-
retarded-minus-half-advanced" potential). This effective potential
was devised by Dirac [10] in order to handle the divergencies
encountered in calculations of the radiation-reaction force in
flat spacetime. We point out, however, that in general the
radiative potential will not be valid in curved spacetime.
Evidently, this potential yields non-causal results: Due to the
scattering off the curvature, the Green's functions of the
massless fields will have a support not only on the light-cone
itself, but also in its interior. Thus, with the advanced
potential, the backreaction force acting on the particle at a
given moment will in general depend on the particle's future
history. Clearly, the radiative potential will inherit this non-
causal behavior, indicating that in general this potential is not
valid in curved spacetime. Although the radiative potential yields
the correct results for $\dot E$   and $\dot L_z$   in Kerr
spacetime [4] , it is not clear whether the corresponding results
for $\dot Q$  will be correct too.

	The inapplicability of the retarded potential in curved
spacetime motivates us to base our approach directly on the
retarded potential. For concreteness and for conceptual clarity,
let us refer at this stage of the discussion to the
electromagnetic radiation reaction (later we shall generalize this
discussion to a scalar field and implement it for explicit
calculations). Consider an electrically-charged particle that
moves in curved spacetime. In principle, the backreaction force
has a simple origin: It is just the Lorentz force, experienced by
the particle due to its interaction with its own electromagnetic
field. However, the field of a point-like particle is divergent at
the particle's location, and any attempt to calculate the
backreaction force must first address this conceptual difficulty.
In flat space, the radiative potential is often used in order to
overcome this difficulty. Unfortunately, this method is not valid
in curved spacetime. A fully-relativistic analysis of
electromagnetic radiation-reaction in curved spacetime was carried
out by DeWitt and Brehme [11]. This analysis is based on the
concept of a world-tube (with finite radius  $\varepsilon $ )
surrounding the charged particle. By taking the limit $\varepsilon
\to 0$ , DeWitt and Brehme obtained a closed expression for the
electromagnetic backreaction force. However, this expression is
not so useful for explicit calculations: It includes a "tail"
term, given by an integral over proper time along the particle's
worldline. It is hard to calculate this tail-integral explicitly -
especially because one does not know the explicit form of the
Green's function that appears in the integrand.

	Our approach to the problem is based on the following
observation: Although the particle's self field is divergent, the
modes of the field (e.g. the spin-weighted spherical harmonics for
Schwarzschild gemetry, and the spin-weighted spheroidal harmonics
for Kerr gemetry) are everywhere finite and well-defined - even at
the limit of a point-like source. Thus, one may simply calculate
the contribution of each mode to the radiative evolution (by a
direct application of the Lorentz-force formula to the interaction
of the particle with the mode in question), and then sum over the
modes. By this decomposition into modes we achieve two goals:
First, we remove the singularity of the self-field.
{\footnote {$^1$}
{The contribution of each mode to the radiation-reaction is
guaranteed to be finite. One still may be concerned about the
convergence of the sum over the modes. At least for $\dot E$ and
$\dot L_z$ , this sum is finite. See the discussion at the end of
the paper.}}
Second, this is the only practical way to solve the field
equations for a Kerr background.

	We still have to address the issue of renormalization. It is
well known, already from flat space, that in general the retarded
backreaction force must be renormalized before it can be used for
explicit radiation-reaction calculations. The same phenomenon
occurs for charged particles in curved spacetime. [11] (Indeed, it
was this complication which motivated Gal'tsov [4] to use the
radiative potential). Fortunately, it turns out that in the
problem considered here no such force-renormalization is needed:
The only need (and justification) for the force-renormalization
originates from the renormalization of the mass-parameter. It
therefore follows that, from its very nature and definition, the
force-renormalization must take the form
$$F_{eff}^\alpha =F_{ret}^\alpha +\mu _{se}a^\alpha\;, \eqno{(1)}$$
where $F_{ret}^\alpha $ is the retarded backreaction force,
$F_{eff}^\alpha $  is the effective (i.e. renormalized) force,
$a^\alpha $  is the covariant four-acceleration, and $\mu _{se}$
is the parameter that renormalizes the particle's mass (the
particle's "self-energy"). This argument [and Eq. (1)] holds in
both flat and curved spacetimes.
{\footnote {$^2$}
{Note the consistency of Eq. (1) with the results obtained in
Ref. [11] for electromagnetic radiation reaction in curved
spacetime; See in particular Eq. (5.23) there.}}
In the problem
discussed here, the radiative evolution of orbits around a Kerr
black hole, no external force is considered, so (within the
framework of the adiabatic approximation) $a^\alpha =0$  and
therefore
$F_{eff}^\alpha =F_{ret}^\alpha $ .
{\footnote {$^3$}
{This statement can be made more precise in terms of the world-
tube approach of Ref. [11] . For any finite radius  $\varepsilon $ ,
the self-energy term in the backreaction force is well-defined
[cf. Eq. (5.23) there] - and vanishes for  $a^\alpha =0$  . One
can then safely take the limit  $\varepsilon \to 0$ .}}

	We turn now to the concrete calculations of the radiative
evolution of generic orbits in Kerr spacetime. For simplicity, we
shall focus here on (massless, minimally-coupled) scalar-field
radiation reaction. The generalization of this calculation scheme
to electromagnetic and gravitational radiation is outlined in Ref.
[9]. Let $C$ denote the constant in question - $E$ , $L_z$ or $Q$ .
All these constants may be expressed explicitly as functions of
coordinates and (covariant) four-velocity, i.e. $C=C(x^\beta
,u_\alpha )$  . Thus, we may write [6]
$$E=-u_t\quad ,\quad L_z=u_\varphi \quad ,\quad Q=u_\theta
^{\;2}+\cos ^2\theta \,[a^2(1-u_t^{\;2})+\sin ^{-2}\theta
\,u_\varphi ^{\;2}]\;\;.$$
(We use Boyer-Lindquist coordinates.) The force induced by a
scalar field $\psi (x^\alpha )$
 on a particle with a scalar charge $q$ is given by
$$F_\alpha =q(\psi _{,\alpha }+u_\alpha u^\beta \psi _{,\,\beta
})\equiv q\psi _{:\,\alpha } \;\;;
 \eqno{(2)}$$
 this is the scalar-field analog of the electromagnetic Lorentz
force [4] . [At this stage,
$\psi (x^\alpha )$  may be thought of as any prescribed scalar
field; We shall soon apply this formalism to the backreaction
problem.] It is not difficult to show [9] that the parameters $C$
evolve according to
$${{dC} \over {d\tau }} =C^\alpha F_\alpha =qC^\alpha \psi
_{:\,\alpha } \;\;, \eqno{(3)}$$
where $\tau $  is the particle's proper time and
$C^\alpha \equiv \mu ^{-1}\,\partial C / \partial u_\alpha $ .

	Next, we decompose $\psi $  into its Teukolsky modes [12] :
$$\psi =\sum\limits_{lm\omega } {\psi ^{lm\omega }}\quad ,\quad
\psi ^{lm\omega }=R_{lm\omega }(r)S_{lm\omega }(\theta
)e^{i(m\varphi -\omega t)} \;\;. \eqno{(4)}$$
 From Eqs. (2,3), the contribution of each mode to the backreaction
force and to $dC / d\tau $  is
$$F_\alpha ^{lm\omega }=q\psi _{:\,\alpha }^{lm\omega }\quad
,\quad \left( {{dC} \over {d\tau }} \right)_{lm\omega }=qC^\alpha
\psi _{:\,\alpha }^{lm\omega } \;\;, \eqno{(5)}$$
where
$$\psi _{:\,\alpha }^{lm\omega }\equiv \psi _{,\,\alpha
}^{lm\omega }+u_\alpha u^\beta \psi _{,\,\beta }^{lm\omega }\;\;.$$
The contribution of each mode $l,m,\omega $  to the change in
$C$ is thus
$$\Delta C_{lm\omega }=q\int {C^\alpha \psi _{:\,\alpha
}^{lm\omega }}\,d\tau\;\;. \eqno{(6)}$$
In this expression, the integrand is to be evaluated along the
particle's worldline $x^\alpha (\tau )$.

	In order to apply Eq. (6) to the backreaction problem, we take
$\psi ^{lm\omega }$  to be the retarded-field modes associated
with the charged particle itself. To determine these modes, one
decomposes the particle's charge-density function $\rho (x^\alpha
)$  into its modes according to [12]
$$4\pi \Sigma \rho =\sum\limits_{lm\omega } {}T_{lm\omega
}(r)S_{lm\omega }(\theta )e^{i(m\varphi -\omega t)}\;\;,\eqno{(7)}$$
where  $\Sigma \equiv r^2+a^2\cos ^2\theta $. The radial
functions $R_{lm\omega }(r)$  are then determined by the Teukolsky
equation [12]
$$(\Delta \,R_{lm\omega })\;+V_{lm\omega }(r)R_{lm\omega }=-
T_{lm\omega }\eqno{(8)}$$
(with retarded boundary conditions), where  $\Delta \equiv r^2-
2Mr+a^2$, a prime denotes $d /dr$, and $V_{lm\omega }(r)$
is some real potential. We take the charge-density function
 $\rho (x^\alpha )$  to be that of a point-like particle:
$$\rho (x^\alpha )=q\sqrt {-g}^{\,-1}\int {\delta \left( {x^\alpha
-x^\alpha (\tau )} \right)d\tau }\;\;,\eqno{(9)}$$
where $x^\alpha (\tau )$  denotes the particle's worldline. (This
approximation demands that the particle's radius $d$ will satisfy
$d<<M$ , which we shall assume.) Note that the thus-obtained
radial functions $R_{lm\omega }(r)$  are everywhere smooth ($C^1$ ).
{\footnote {$^4$}
{The special case of circular orbits (r=const) is exceptional,
as the radial functions are not  $C^1$  at the particle's
location. This simpler case has to be treated separately.}}

	Finally, the long-term evolution rate (in terms of asymptotic
time $t$) is given by
$$\dot C=\sum\limits_{lm\omega } {}\dot C_{lm\omega }\quad ,\quad
\dot C_{lm\omega }=\Delta t^{-1} Real\left( {\Delta C_{lm\omega }}
\right)\;\;.\eqno{(10)}$$
(The imaginary part of  $\Delta C_{lm\omega }$  will cancel out
upon summation, because  $\Delta C_{l,-m,-\omega }=\Delta
C_{lm\omega }^*$  [9].) Here,  $\Delta t\equiv t(\tau _2)-t(\tau
_1)$   , where $\tau _1$  and  $\tau _2$  are the lower and upper
integration limits in Eq. (6). For strictly periodic orbits, we
choose $\tau _1$  and  $\tau _2$  such that $\Delta \tau \equiv
\tau _2-\tau _1$  is the orbital period; For the generic orbits in
Kerr spacetime, which are not strictly periodic but rather quasi-
periodic, we take  $\Delta \tau $  to be sufficiently large, so
that the quasi-periodicity will yield a well-defined limit in Eq.
(10) (formally speaking, for quasi-periodic orbits we shall be
interested in the limit $\Delta \tau \to \infty $ ).

	Equations (6-10) constitute a closed scheme for the computation
of the (scalar-field) radiative evolution of all constants of the
motion in Kerr spacetime. The calculation may be further
simplified, however, by using Eq. (9) to rewrite Eq. (6) as
$$\Delta C_{lm\omega }=\int {C^\alpha \psi _{:\,\alpha }^{lm\omega
}}\,\,\Sigma \rho \,\sin \theta \,d^4x^\alpha \eqno{(11)}$$
(recall  $\sqrt {-g}=\Sigma \sin \theta $  ).
{\footnote {$^5$}
{$C^\alpha $  is only defined along the particle's worldline;
However, the integrand in Eq. (11) vanishes elsewhere anyway.}}
We shall now treat the cases $C=E$ and  $C=L_z$  , which are
especially simple. Let  $P$ denote either $E$ or $L_z$  , and let
$p$ denote  $\omega $  or m , correspondingly. From Eq. (4) and
the definitions of  $C^\alpha $  and $C$ one then finds (ignoring
terms which just oscillate in time)
$$\Delta P_{lm\omega }={{ip} \over \mu }\int {\psi ^{lm\omega
}}\,\,\Sigma \rho \,\sin \theta \,d^4x^\alpha\;\;. \eqno{(12)}$$
Finally, using the decompositions (4,7), and the orthogonality of
the modes, one obtains
$$\dot P_{lm\omega }={{ip} \over {4\mu} }\int {R_{}^{lm\omega
}}\,(r)T_{lm\omega }^*\,(r)dr+C.C. \;\;. \eqno{(13)}$$

	It would be interesting to compare Eq. (13) to the standard
expressions for energy and angular-momentum radiative losses. By
virtue of Eq. (8), it is possible to integrate Eq. (13)
explicitly:
$$\dot P_{lm\omega }={{ip} \over {4\mu} }\left[ {\Delta
\,R_{}^{*lm\omega }\,R_{,r}^{lm\omega }}
\right]_{r_+}^\infty +C.C. \;\;, \eqno{(14)}$$
where  $r_+$  denotes the horizon's radius (this equation may
easily be checked by a direct differentiation). Taking into
account the asymptotic behavior of the radial functions in both
asymptotic limits [7] , it is now straightforward to show that Eq.
(14) is exactly of the desired form; Namely,  $\dot P_{lm\omega }$
is the sum of two terms - the fluxes (per unit mass  $\mu $ ) of
energy or azimuthal angular momentum that flow (i) to infinity and
(ii) into the horizon -  exactly as calculated by the standard
method [7] .

	The more interesting application of our scheme would be the
calculation of $\dot Q$  (here there are no previous results to
compare with). Equations (10-11) provide a closed scheme for the
calculation of $\dot Q$ . In Ref. [9], however, we derive from
Eqs. (10-11) a more explicit expression for  $\dot Q_{lm\omega }$ :
$$\dot Q_{lm\omega }=\mu ^{-1}\int {\left( {HR_{}^{lm\omega
}\,T_{lm\omega }^*\,-R_{}^{lm\omega }\,\tilde T_{lm\omega }^*\,-
R_{,r}^{lm\omega }\hat T_{lm\omega }^*\,\,} \right)}_{}^{}dr+C.C.\;\;,
\eqno{(15)}$$
where
$$H\equiv ir\Delta ^{-1}\left[ {\omega E(r^3+ra^2+2Ma^2)-2(\omega
L_z+mE)Ma+mL_z(2M-r)} \right]_{}$$
and  $\tilde T_{lm\omega }(r)$  and  $\hat T_{lm\omega }\,(r)$
are functions analogous to $T_{lm\omega }\,$  , obtained by
replacing the factor  $\rho $  at the left-hand side of Eq. (7)
with  $\tilde \rho \equiv 2ru^r\rho $   or   $\hat \rho \equiv
\Delta u_{r\,}\rho $  , correspondingly. Equation (15) thus allows
a straightforward numerical calculation of  $\dot Q_{lm\omega }$
(and $\dot Q$  ).

	We wish to emphasize that no attempt was made here to provide a
complete and fully-rigorous treatment of all the subtle issues
involved in the radiation-reaction problem. Rather, the goal here
was to propose a simple and straightforward approach to the
problem, which appears to yield promising results: It recovers the
standard results for $\dot E$   and $\dot L_z$   in Kerr spacetime
(at least for a scalar field) - and also yields an explicit
expression for $\dot Q$  . Our approach is based on the delta-
function approximation, Eq. (9). This approximation has
extensively been used over the years for radiation-reaction
calculations (see for example Refs.[3-5] and references therein).
We believe that this approximation is justified in our case. We
hope that in the future it will be possible to apply more rigorous
methods for the calculation of the radiative evolution of orbits
in Kerr spacetime.

	One may be concerned about the convergence of the sum over the
modes in Eq. [10] . This series is known to converge for $C=E$ and
$C=L_z$  , and this makes us optimistic about its convergence for
$C=Q$ as well. The explicit calculation scheme presented here
allows one to verify this convergence (for $Q=C$) numerically -
and perhaps even analytically.

	Finally, we mention several tests which may be applied to our
result for $\dot Q$ , Eq. (15). The first such test would be, of
course, to verify converges of the sum over the modes (see above).
Another simple test can be applied to the special case a=0 (the
Schwarzschild case): Due to the spherical symmetry of the latter,
the orbital plane must be preserved. A simple calculation then
yields  $\dot Q=2(Q / L_z)\,\dot L_z$  . Although this test is
limited to the case a=0 , it is conceivable that if our scheme is
faulty, it will yield wrong results in that case too. Finally, for
equatorial orbits ($Q=0$) in Kerr spacetime, $\dot Q$  must vanish
identically.

	I would like to thank Eric Poisson, Sam Finn, Eanna Flanagan
and Kip Thorne for stimulating discussions and helpful comments.

\vskip 1cm
\leftline {\bf References}
\vskip .5cm
\item {[1]}
K. Danzmann at al. , LISA: Proposal for a laser-
interferometric gravitational wave detector in space, Max-Planck
Institute fur Quantenoptik Report 177, May 1993 (unpublished).
\item {[2]} C. Cutler at al. Phys. Rev. Lett.
  {\bf 70} , 2984 (1993).
\item {[3]} See the discussion in C. Cutler,
 D. Kennefick and E. Poisson,
Phys. Rev. D. {\bf 50}, 3816  (1994).
\item {[4]} D. V. Gal'tsov,
 J. Phys. A {\bf 15} 3737 (1982) .
\item {[5]} M. Shibata, Prog. Theor. Phys. {\bf 90} , 595 (1993).
\item {[6]} See e.g. C. W. Misner, K. S. Thorne and J. A. Wheeler,
Gravitation (Freeman, San Francisco, 1973) ,  \S 33.5 .
\item {[7]} S. A. Teukolsky and W. H.
Press, Astrophys. J. {\bf 193} 443 (1974).
\item {[8]} See however the approximate method proposed
 in Ref. [5] .
\item {[9]} A. Ori, "Radiative evolution of
 the Carter constant for orbits
in Kerr", preprint (to be published).
\item {[10]} P. A. M. Dirac, Proc. Roy. Soc. {\bf A167}, 148 (1938).
\item {[11]} B. S. DeWitt and R. W. Brehme,
 Ann. Phys. (U.S.A.) {\bf 9} 220
(1960) .
\item {[12]} S. A. Teukolsky, Astrophys. J. {\bf 185}, 635 (1973).

\bye